%% file: fullpaper.tex
\documentclass[conference,a4paper]{IEEEtran}
\IEEEoverridecommandlockouts
\usepackage{cite}
\usepackage{amsmath,amssymb,amsfonts}
\usepackage{algorithm}
\usepackage{algorithmicx}
\usepackage{algpseudocode}
\usepackage{graphicx}
\usepackage{booktabs}
\usepackage{textcomp}
\usepackage{xcolor}
\usepackage{hyperref}
\usepackage{comment}
\usepackage[nolist]{acronym}
\usepackage{paralist}
\usepackage[skip=0.333\baselineskip]{caption}
\usepackage{subcaption}
\def\BibTeX{{\rm B\kern-.05em{\sc i\kern-.025em b}\kern-.08em
    T\kern-.1667em\lower.7ex\hbox{E}\kern-.125emX}}

\begin{document}
\makeatletter
\def\ps@IEEEtitlepagestyle{%
\def\@oddfoot{\parbox{\textwidth}{\centering\normalsize{979-8-3503-9042-1/24/\$31.000 \copyright 2024 IEEE}\vspace{2em}}
}%
}

\input{acronyms}

\bstctlcite{IEEEexample:BSTcontrol}

\title{
Encryption-Aware Anomaly Detection in Power Grid Communication Networks
}

\author{
\IEEEauthorblockN{%
Ömer Sen\IEEEauthorrefmark{1}\IEEEauthorrefmark{2},
Mehdi Akbari Gurabi\IEEEauthorrefmark{1}\IEEEauthorrefmark{2},
Milan Deruelle\IEEEauthorrefmark{1},
Andreas Ulbig\IEEEauthorrefmark{1}\IEEEauthorrefmark{2},
Stefan Decker\IEEEauthorrefmark{1}\IEEEauthorrefmark{2},
}

\IEEEauthorblockA{%
\IEEEauthorrefmark{1}\textit{RWTH Aachen University,} Aachen, Germany |
\IEEEauthorrefmark{2}\textit{Fraunhofer FIT,} Aachen, Germany\\
Email: \{oemer.sen, mehdi.akbari.gurabi, andreas.ulbig, stefan.decker\}@fit.fraunhofer.de} milan.deruelle@rwth-aachen.de
}

\maketitle

\begin{abstract}
The shift to smart grids has made electrical power systems more vulnerable to sophisticated cyber threats. To protect these systems, holistic security measures that encompass preventive, detective, and reactive components are required, even with encrypted data. However, traditional intrusion detection methods struggle with encrypted traffic, our research focuses on the low-level communication layers of encrypted power grid systems to identify irregular patterns using statistics and machine learning. Our results indicate that a harmonic security concept based on encrypted traffic and anomaly detection is promising for smart grid security; however, further research is necessary to improve detection accuracy.
\end{abstract}


\input{chapter1}
\input{chapter3}
\input{chapter4}

\input{chapter5}

\section*{Acknowledgment}
\begin{minipage}{0.75\columnwidth}%
This work is funded by BMBF (03SF0694A, Beautiful) and the EU H2020 R\&I programme (101020560, CyberSEAS).
\end{minipage}
\hspace{0.02\columnwidth}
\begin{minipage}{0.15\columnwidth}%
	\includegraphics[width=\textwidth]{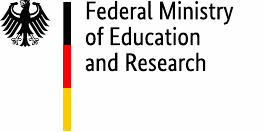}
        \includegraphics[width=\textwidth]{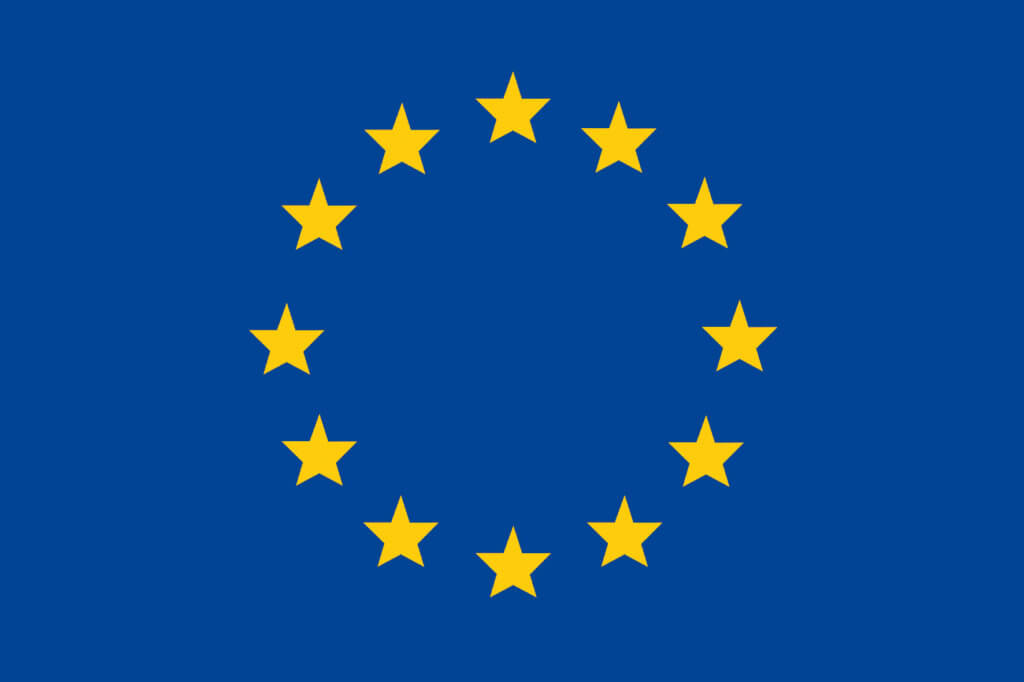}
\end{minipage}
\vspace{-0.5em}

\bibliographystyle{IEEEtran}
\bibliography{fullpaper}

\end{document}

%% file: acronyms.tex
\acrodef{DSO}{Distribution System operator}
\acrodef{DBSCAN}{Density-Based Spatial Clustering of Applications with Noise}
\acrodef{HDBSCAN}{Hierarchical Density-Based Spatial Clustering of Applications with Noise}
\acrodef{TCP}{Transmission Control Protocol}
\acrodef{TTL}{Time to Live}
\acrodef{IQR}{Interquartile Range}
\acrodef{MTU}{Master Terminal Unit}
\acrodef{VRTU}{Virtual Remote Terminal Unit}
\acrodef{DBCV}{Density-Based Clustering Validation}
\acrodef{DoS}{Denial of Service}
\acrodef{SSL}{Secure Sockets Layer}
\acrodef{IEC104}{IEC 60870-5-104}
\acrodef{TLS}{Transport Layer Security}
\acrodef{SCADA}{Supervisory Control and Data Acquisition}
\acrodef{DER}{Distributed Energy Resources}
\acrodef{ICT}{Information and Communication Technologies}
\acrodef{CIA}{Confidentiality, Integrity and Availability}
\acrodef{ICS}{Industrial Control System}
\acrodef{IT}{Information Technology}
\acrodef{OT}{Operational Technology}
\acrodef{MulVAL}{Multi-host, Multi-stage Vulnerability Analysis}
\acrodef{NVD}{National Vulnerability Database}
\acrodef{OVAL}{Open Vulnerability and Assessment Language}
\acrodef{TTC}{Time-to-Compromise}
\acrodef{IDS}{Intrusion Detection System}
\acrodef{CVE}{Common Vulnerabilities and Exposures}
\acrodef{CVSS}{Common Vulnerability Scoring System}
\acrodef{Ac}{Access Complexity}
\acrodef{Au}{Authentication}
\acrodef{Ex}{Exploitability}
\acrodef{CPT}{Cyber-Physical Digital Twin}
\acrodef{DT}{Decision Tree}
\acrodef{GAN}{Generative Adversarial Network}
\acrodef{RF}{Random Forest}
\acrodef{SVM}{Support Vector Machine}
\acrodef{MCC}{Matthews correlation coefficient}
\acrodef{AUC}{Area Under Curve}
\acrodef{ROC}{Receiver Operating Characteristic}
\acrodef{TP}{True Positive}
\acrodef{TN}{True Negative}
\acrodef{FP}{False Positive}
\acrodef{FN}{False Negative}
\acrodef{CNB}{Complement Naïve Bayes}
\acrodef{XGB}{Extreme Gradient Boosting}
\acrodef{MQTT}{Message Queuing Telemetry Transport}
\acrodef{TTC}{Time-to-Compromise}
\acrodef{ML}{Machine Learning}
\acrodef{HMI}{Human-Machine-Interface}
\acrodef{CI}{Confidence Interval}

%% file: chapter1.tex
%
\section{Introduction}\label{sec:intro}

As power grids increasingly integrate \ac{ICT}, they become more interconnected with external systems. This reduces isolation barriers and introduces new cybersecurity risks~\cite{krause2021cybersecurity}. A prominent instance is the 2015 Ukraine cyberattack~\cite{case2016analysis}, which led to a large-scale blackout. Ensuring cybersecurity in power systems is an ongoing process, focusing on prevention, detection, and response, guided by standards such as IEC 62351~\cite{cleveland2012iec} and IEC 62443~\cite{piggin2013development}. Moreover, emerging cybersecurity regulations like NIS2~\cite{Coppolino2023} emphasize the importance of enhancing the resilience and protection of critical infrastructures from a legal standpoint.

Vulnerabilities are particularly evident in protocols such as \ac{IEC104}, which are widely used in \ac{SCADA} systems but lack fundamental security features such as authentication and encryption~\cite{lin2018understanding, radoglou2019attacking}. Enhancing security by incorporating measures such as \ac{TLS} encryption and \ac{IDS}, in compliance with regulations like Germany's IT-Sicherheitsgesetz 2.0, is challenging due to the necessity of intercepting data for detection while simultaneously protecting sensitive information~\cite{brzostek2022germany}.

Anomaly detection approaches analyze network traffic, comparing observed data against models of normal behavior. These methods, often utilizing \ac{ML}, require process data for both training and operational execution. Middle-boxes used for \ac{IDS} implementation can weaken encryption~\cite{sherry2015blindbox}. A potential alternative is an \ac{IDS} that functions independently of higher communication layers, thus remaining compatible with encrypted traffic, as described in IEC 62351. However, its performance in \ac{SCADA} systems within power grids remains unproven.

Research on \ac{IDS} for encrypted communications has primarily focused on anomaly-based systems designed to detect protocol misuse, decrypt data for payload analysis, or utilize network flow data to address the privacy and security challenges posed by decryption~\cite{kovanen2016survey}. Comprehensive surveys have reviewed a variety of approaches, from feature selection techniques to complete system implementations~\cite{papadogiannaki2021survey}. While most of these studies apply anomaly detection using statistical analysis and machine learning, some propose signature-based detection methods~\cite{mirsky2018kitsune}.

Machine learning has been extensively studied for classifying encrypted malware traffic, with Random Forests demonstrating the importance of feature engineering~\cite{zolotukhin2015data}. In high-speed environments, real-time unsupervised \ac{IDS} using \ac{DBSCAN} clustering has been developed to identify potential threats~\cite{amoli2013real}. Autoencoders have also been employed for online intrusion detection, focusing on network traffic monitoring, feature extraction, and behavior modeling~\cite{anderson2017machine}.

A hybrid approach that combines supervised learning algorithms with \ac{HDBSCAN} for unclassified traffic has proven effective in distinguishing between benign and malicious traffic, thus improving the detection of unknown or zero-day attacks~\cite{shahbandayeva2022network}. The effectiveness of \ac{HDBSCAN} in various anomaly detection applications has been further demonstrated~\cite{ahasan2022benchmarking}. Additionally, a data mining technique using \ac{SSL}/\ac{TLS} and \ac{DBSCAN} has been proposed for detecting \ac{DoS} attacks, with a focus on identifying anomalies while maintaining low false positive rates~\cite{papadogiannaki2021acceleration}. This technique segments traffic and models normal behavior, flagging outliers as potential intrusions.

This paper addresses the gap in comparative studies on encryption-informed \ac{IDS} methodologies in smart grids, particularly concerning legacy systems and compliance with IEC 62351. It explores anomaly detection in lower-level, non-encrypted communication layers, analyzing challenges in encrypted \ac{ICS} communications, and evaluates flow-based anomaly detection systems to assess the impact of encryption on \ac{IDS} performance.

%% file: chapter3.tex
\section{Encryption-Aware IDS}\label{sec:method}

\subsection{Approach}\label{subsec:method_overview}

\begin{figure}[htbp]
\centerline{\includegraphics[width=\columnwidth]{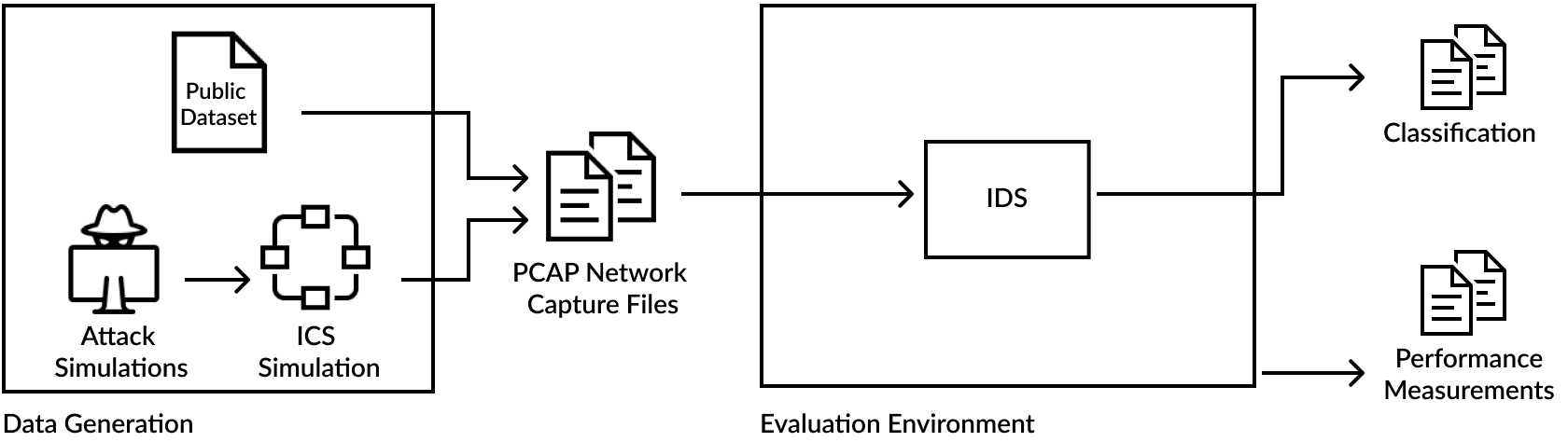}}
\caption{Visualization of the conceptual approach, illustrating the three main components of our system: Data Generation, the IDS, and the Evaluation Environment.}
\label{fig:method_overview}
\vspace{-0.5em}
\end{figure}

Figure~\ref{fig:method_overview} presents our approach for evaluating the transferability of encrypted \ac{IDS} methodologies to \ac{ICS} communication, focusing on the widely used \ac{IEC104} protocol in power grids. We utilize TCP/IP metadata such as endpoint addresses, data sequence order, and connection-oriented information for anomaly detection and gather datasets from public sources and data generation using a \ac{CPT}. These datasets are used to evaluate \ac{SSL}-based anomaly detection algorithms, assessing their effectiveness in identifying anomalies and attacks within encrypted \ac{ICS} communications.

The primary challenges we face are obtaining suitable data, performing anomaly detection, and evaluating the \acp{IDS}. We aim to cover a wide range of attacks and anomalies for an effective evaluation of zero-day attack detection performance. Although public datasets for unencrypted \ac{IEC104} communications exist, such as~\cite{matouvsek2022ics}.They include four attack datasets (Report Block, Replay, Value Change, and Masquerading) and one normal traffic dataset. However, none of these datasets are available for communications protected by IEC TS 62351-3. Utilizing the \ac{CPT}, as developed in our previous work~\cite{sen2023digital}, we generate datasets of both encrypted and unencrypted \ac{IEC104} communications, with adaptations to secure communication per IEC TS 62351-3 using \ac{TLS}. These scenarios are important because they represent the early stages of multi-stage attacks, providing valuable insight into system vulnerabilities before further escalation. Additionally, they test the effectiveness of intrusion detection systems (IDS) in identifying subtle anomalies during reconnaissance. These data are described in Table~\ref{table:generated_datasets}. This setup allows us to:
\begin{enumerate}
    \item Evaluate algorithm performance on encrypted (per IEC TS 62351-3) and unencrypted \ac{IEC104} communication.
    \item Investigate the use of metadata available in both types of communications to estimate the performance of an IDS in encrypted environments.
\end{enumerate}

\begin{table}[htbp]
\centering
\caption{Generated dataset identifiers per anomaly}
\label{table:generated_datasets}
\begin{tabular}{|l|p{4cm}|l|l|}
\hline
\textbf{Anom.} & \textbf{Scenarios} & \textbf{No TLS} & \textbf{TLS} \\
\hline
AN1 & Standard Operation (Benign Data) & D2-1-1 & D2-2-1 \\
AN2 & 2-hop Targeted Manipulation 1 & D2-1-2 & D2-2-2 \\
AN3 & 2-hop Targeted Manipulation 2 & D2-1-3 & D2-2-3 \\
AN4 & 2-hop vRTU Slowdown & D2-1-4 & D2-2-4 \\
AN5 & 2-hop vRTU Shutdown & D2-1-5 & D2-2-5 \\
AN6 & Telnet Data Exfiltration & D2-1-6 & D2-2-6 \\
AN7.1 & Reconnaissance - Default Options & D2-1-7.1 & D2-2-7.1 \\
AN7.2 & Reconnaissance - No ARP or ND & D2-1-7.2 & D2-2-7.2 \\
AN7.3 & Reconnaissance - TCP Connect & D2-1-7.3 & D2-2-7.3 \\
AN7.4 & Reconnaissance - TCP SYN Scan & D2-1-7.4 & D2-2-7.4 \\
AN7.5 & Reconnaissance - TCP NULL Scan & D2-1-7.5 & D2-2-7.5 \\
AN7.6 & Reconnaissance - TCP FIN Scan & D2-1-7.6 & D2-2-7.6 \\
AN7.7 & Reconnaissance - Xmas Scan & D2-1-7.7 & D2-2-7.7 \\
\hline
\end{tabular}
\vspace{-1.5em}
\end{table}

\begin{figure}[htbp]
\centerline{\includegraphics[width=\columnwidth]{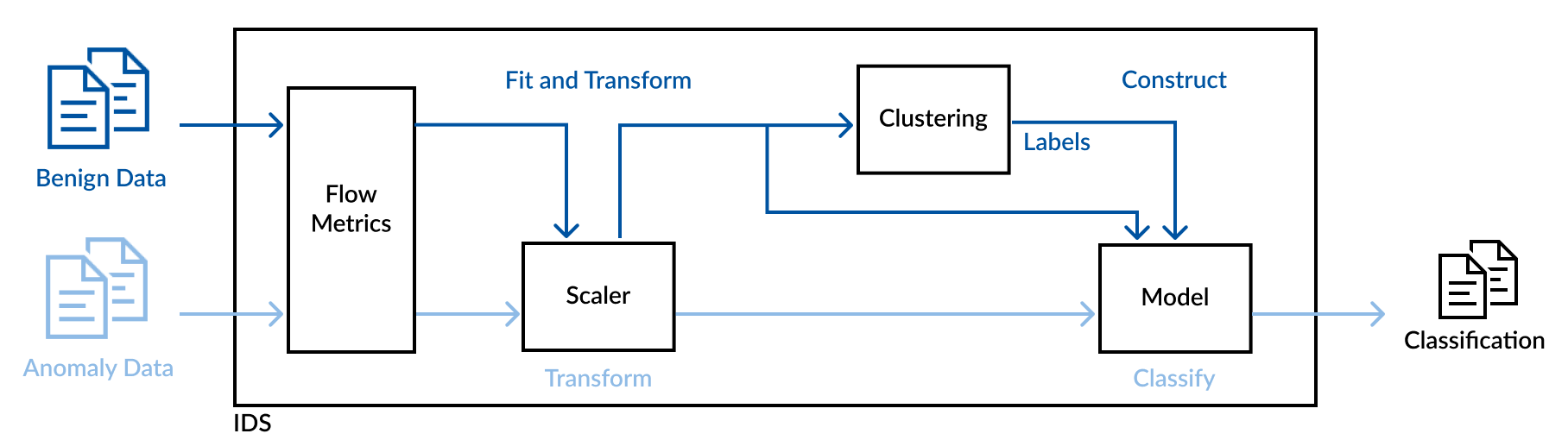}}
\caption{The conceptual approach of the IDS structure, showcasing the main components of the IDS.}
\label{fig:method_ids}
\vspace{-1em}
\end{figure}

The proposed approach consists of the following steps:
\begin{enumerate}
    \item Flow Metrics are based on packets exchanged in a session, calculated using either a sliding window or time-slot approach.
    \item Metrics such as Inter Packet Time, Packet Size, and \ac{TTL} are scaled using \ac{IQR} to ensure robust handling of outliers.
    \item Anomaly detection uses \ac{DBSCAN} and \ac{HDBSCAN}, classifying flows based on their distance to the nearest cluster.
\end{enumerate}

We adopt the flow definition by~\cite{draper2016characterization}, encompassing all packets transmitted between a sender and receiver in a session (cf. Figure~\ref{fig:method_ids}). We categorize flows by their unique combination of IP addresses, port numbers, and protocol such as TCP/IP stacks, determining a sender either as the connection initiator or from the first observed packet in an ongoing session. However, considering the structure of \ac{ICS} networks and peculiarities like many short-lived flows or limited numbers of FIN/RST packets, we propose two methods for calculating flow metrics: a Windowed Approach and a Slotted Approach. These methods respectively utilize a sliding window or divide continuous flows into time slots for metric calculations. The Windowed Approach computes for a packet $p$, its corresponding flow
$f$, a window size time $t$ and the timestamp $ts$ as follows:
\begin{equation}
\begin{aligned}
    window(f, p, t) = \{x \mid x \in f \land 0 \leq p_{\text{ts}} - x_{\text{ts}} \leq t\}
\end{aligned}
\end{equation}
The Slotted Approach divides flows into time slots, terminated by either a FIN/RST packet or exceeding the slot size \( t_{\text{slot}} \).

Upon grouping packets into flows, various metrics are calculated, such as Inter Packet Time, percentages of different flag types (SYN, ACK, PSH, RST, FIN), Packet Size, \ac{TTL}, and Window Size. Each metric's mean, maximum, and minimum values are computed (except for flag percentages).

Given the diverse scales of calculated metrics, robust scaling is employed to maintain the range of the majority of data points while preserving the distribution and distances between points. Robust scaling is applied using \ac{IQR}:
\begin{equation}
    X_{\text{scaled}} = \frac{X - X_{\text{median}}}{X_{\text{\ac{IQR}}}}
\end{equation}
This scaling ensures a balanced representation of outliers and the majority of data points and scales based on the \ac{IQR}, effectively balancing data point range and outlier distances.

We incorporate various approaches, including ~\cite{mirsky2018kitsune}, and \ac{DBSCAN} and \ac{HDBSCAN}-based methods \cite{zolotukhin2015data}. Our model combines these methods, particularly adapting semi-supervised learning approach with \ac{DBSCAN} and \ac{HDBSCAN} clustering algorithms. The detection model is described in Algorithm~\ref{algo:method_detection}. The model classifies flows based on their distance to the nearest cluster in the training data, using maximum pairwise distance $mpdi$ as a threshold for anomaly determination~\cite{zolotukhin2015data}: 
\begin{equation}
    mpdi = \max_{p,q \in C_i} d(p, q)
\end{equation}
\vspace{-1em}

\begin{algorithm}
\caption{Detection Model}
\label{algo:method_detection}
\begin{algorithmic}[1]
\State \textbf{on} Initialisation $\langle training~data, labels\rangle$ \textbf{do:}
\State \quad Compute distance $mpdi$ of each cluster $ci$
\State \quad Initialize nearest neighbor search with training data
\State \textbf{on} receiving flow $\langle f\rangle$ \textbf{do:}
\State \quad Find nearest neighbor $n$ of $f$ in training data
\State \quad Find cluster $ci$ of $n$
\State \quad Compute distance $d$ of $n$ and $f$
\State \quad \textbf{if} $d \leq mpdi$ \textbf{then}
\State \quad \hspace{\algorithmicindent} Classify $f$ as Benign
\State \quad \textbf{else}
\State \quad \hspace{\algorithmicindent} Classify $f$ as Anomaly
\State \quad \textbf{end if}
\end{algorithmic}
\end{algorithm}
\vspace{-1em}

Selecting parameters for \ac{DBSCAN} involves methods like grid-search with Silhouette Score~\cite{rousseeuw1987silhouettes} or \ac{DBCV}~\cite{moulavi2014density}, and setting parameters based on mean pairwise distances of training data points. While k-distance graphs provide a visual parameter range, a combination of hyper-parameter tuning with clustering evaluation metrics is employed for replicable results. Different parameter determination methods will be compared, including Gridsearch with Silhouette Score or \ac{DBCV}, and Mean Pairwise Distance approach.

\subsection{Implementation} \label{sec:implementation}

In this section, we present the realization and implementation of our investigation environment. 
The system architecture comprises two main parts
: data procurement and the \ac{IDS} itself. Data procurement includes publicly available datasets and self-generated datasets, while the \ac{IDS} is divided into two software packages: flowmetrics and evaluation.

For encrypted data evaluation, datasets were generated using a \ac{CPT}~\cite{sen2023digital}, modified for compatibility with IEC TS 62351-3. The \ac{CPT}’s two submodules, \ac{MTU} and \ac{VRTU}, were updated to integrate \ac{TLS} features. A bash script was developed for automated and consistent data generation.

Our flowmetrics package, designed for extracting flow-based statistics from packets, is based on the principles of CICFlowMeter~\cite{ali2022ddos} with additional features like sliding window calculation, slotted calculation, and flow completion callbacks. This package manages the flow of packets using a hierarchy of levels: top-level for flow keys, FlowCollection for each key, and Flow objects for packets management. The packet flow management is implemented using the Algorithm~\ref{algo:method_flowmanagement}.

\begin{algorithm}
\caption{Packet Flow Management}
\label{algo:method_flowmanagement}
\begin{algorithmic}[1]

\algblockdefx[NAME]{ON}{END}%
    [2][Unknown]{\textbf{on} #1 $<$#2$>$ \textbf{do}:}

\State \textbf{on} receiving packet $<p>$ \textbf{do:}
\State \quad $k \gets$ key($p$)
\State \quad \textbf{if} FlowCollection $f$ for key $k$ exists \textbf{then}
    \State \quad \quad \textbf{if} Most recent Flow $q$ of $f$ is not terminated \textbf{then}
        \State \quad \quad \quad Add $p$ to $q$
        \State \quad \quad \quad \textbf{if} $p$ terminates $q$ \textbf{then}
            \State \quad \quad \quad \quad Mark $q$ as terminated
        \State \quad \quad \quad \textbf{end if}
    \State \quad \quad \textbf{else}
        \State \quad \quad \quad  Initialize new Flow with $p$ and add to $f$
    \State \quad \quad \textbf{end if}
\State \quad \textbf{else}
    \State \quad \quad Initialize new FlowCollection with $p$ for key $k$
\State \quad \textbf{end if}

\end{algorithmic}
\end{algorithm}

The evaluation package includes modules for scaling, clustering, model construction, and various helper functions. It adopts scikit-learn's API design for modularity~\cite{hackeling2017mastering}, facilitating the combination and replacement of components. Clustering uses \ac{DBSCAN} and \ac{HDBSCAN} from scikit-learn, and hyper-parameter tuning is achieved through Ray Tune. The model utilizes NumPy and scikit-learn’s KDTree for nearest neighbor searches and classification based on maximum pairwise distance. Preprocessing is conducted using scikit-learn’s RobustScaler. Model construction involves nearest neighbor search and classification based on maximum pairwise distance (cf. algorithm ~\ref{algo:method_detection}), where $mpdi$ is the maximum pairwise distance for the cluster of the nearest neighbor.

%% file: chapter4.tex
\vspace{-0.5em}
\section{Result} \label{sec:result}
In this section, we analyze the performance of an encryption-aware \ac{IDS}, focusing on transforming datasets, labeling complexities, and evaluation metrics.

We assessed an encryption-aware \ac{IDS} by processing datasets into feature vectors with timespan parameters of 10s, 30s, and 60s, using flowmetrics, and aggregating data in a Pandas DataFrame for statistical analysis. 
Particularly during attack phases, the labeling process was meticulously carried out. We distinctly labeled normal data, attack vectors, actual attacks, and their consequent effects to enhance the accuracy of our analysis. Consequently, the feature vectors we derived from this process consisted of 34 dimensions. We measured classification effectiveness with F1-Score, Precision, and Recall, considering labeling constraints, allowing for a nuanced assessment of the model’s performance. Moreover, computational efficiency was evaluated by comparing time and memory usage in slotted and windowed processing modes. 

\subsubsection{Computational Efficiency} \label{subsubsec:result_experiment_comp}

\begin{figure}[hbt!]
\centering
\includegraphics[width=\linewidth]{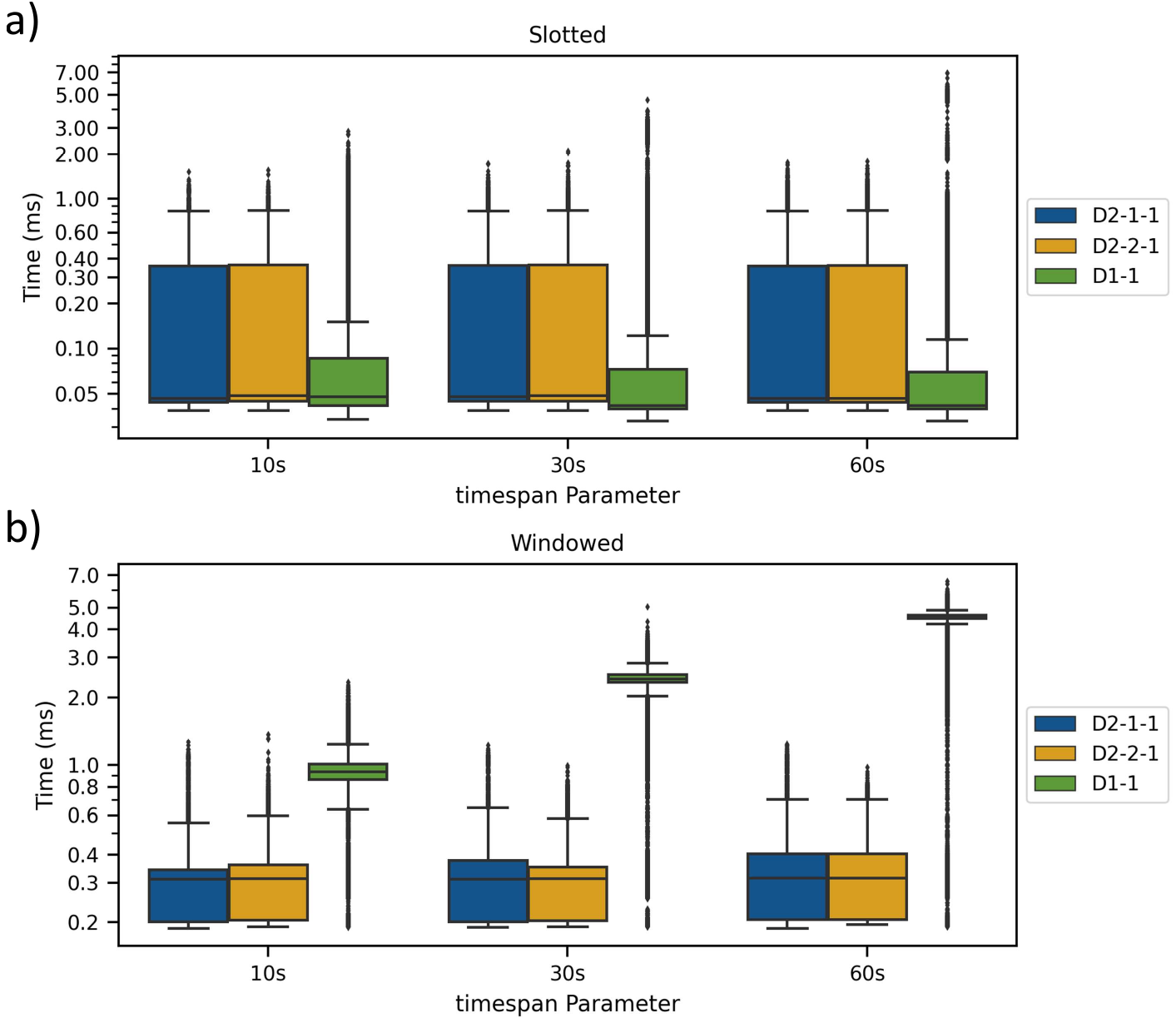}
\caption{Execution time for the slotted (a) and windowed (b) packet processing methods of the flowmetrics component across datasets D2-1-1 (TLS-disabled), D2-2-1 (TLS-enabled), and the normal dataset from D1 (D1-1) by timespan.}
\label{fig:results_comp1a}
\vspace{-0.5em}
\end{figure}

\begin{figure}[hbt!]
\includegraphics[width=\linewidth]{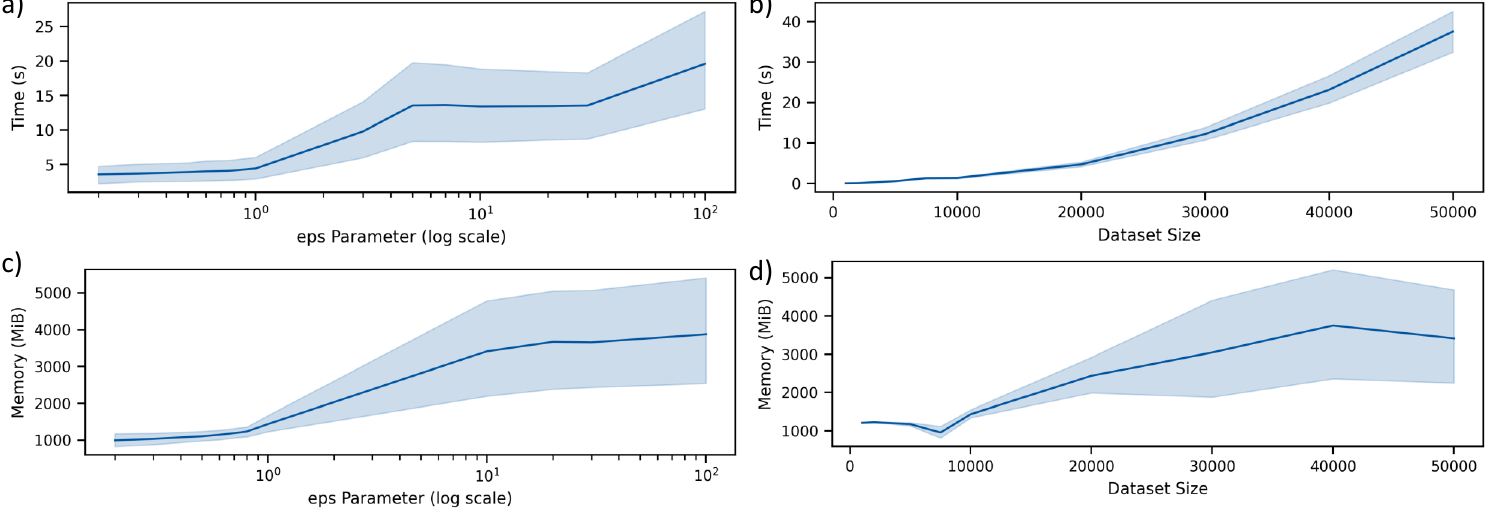}
\caption{ \ac{DBSCAN} clustering time (a,b) and memory usage (c,d) for different $eps$ parameter values and different dataset sizes.}
\label{fig:results_comp2}
\vspace{-1em}
\end{figure}

This section assesses the computational efficiency of our system's components in \ac{ICS} environments, focusing on speed, resource use, scalability, and adaptability, quantified through execution time and memory usage analysis.

Our analysis of execution times in slotted and windowed modes, as shown in Figure~\ref{fig:results_comp1a}, reveals significant performance differences influenced by mode implementation, network structures, and traffic patterns, although neither mode presented a bottleneck in runtime efficiency. The assessment of scoring functions using the normal dataset from D1 in windowed mode highlighted increased execution times and memory usage for larger datasets, 
with challenges in handling large sample sizes leading to system crashes and practical issues. Among the scoring methods, the \ac{DBCV} score, when paired with \ac{HDBSCAN}, proved computationally efficient for evaluating clustering quality.

In evaluating the computational efficiency of \ac{DBSCAN} using the D2-1-1 dataset, it was found that both the eps parameter (permitted variation for cluster boundary) and dataset size significantly affect clustering time and memory usage, as illustrated in Figures~\ref{fig:results_comp2}. Similarly, for \ac{HDBSCAN} using the normal dataset from D1, the parameter for cluster size and dataset size impacted clustering times, with a notable non-linear increase for larger datasets. The construction and classification phases of our model, also tested on the same dataset, demonstrated a non-linear increase in construction time and a linear increase in memory usage for larger datasets, while maintaining stable computational demand during classification across various batch sizes.

\subsubsection{Hyperparameter Tuning} \label{subsubsec:result_experiment_tuning}

\begin{figure}[hbt!]
\centering
\includegraphics[width=\linewidth]{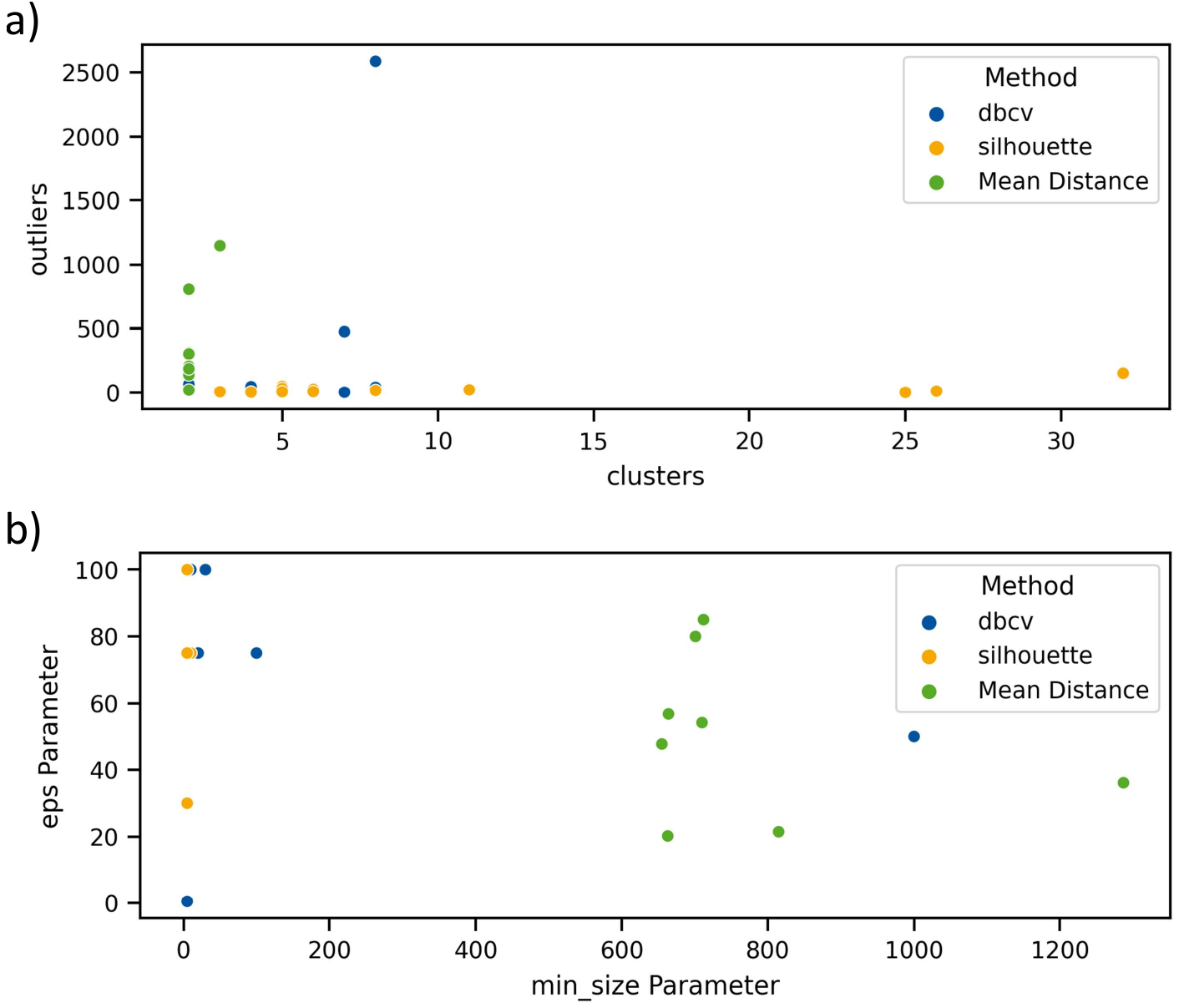}
\caption{Scatterplots illustrating the number of outliers and clusters, and the selected parameters (eps and min\_samples) for different scoring methods.}
\label{fig:results_tuning}
\vspace{-1em}
\end{figure}

Our hyperparameter tuning approach employed our library to explore optimal configurations for clustering algorithms systematically. For \ac{DBSCAN}, k-distance graphs approximated the epsilon value range. In contrast, \ac{HDBSCAN}'s setup encompassed a broader parameter range. 

The tuning process encountered memory constraints in slotted mode, with up to 180 GB needed for some model training scenarios. Figure~\ref{fig:results_tuning} summarizes the tuning results, with the \ac{DBCV} metric showing the highest scores for a 60s timespan. Plot a) illustrates parameter selections, highlighting trends in min\_samples and eps values among the methodologies. In contrast, plot b) displays the variance in the number of clusters and outliers across different configurations.

Our findings highlight a limitation in the windowed mode, where feature vector generation exceeded the scalability of clustering and scoring algorithms. The Silhouette score preference for fewer, larger clusters diverging from the k-distance graph suggested range, impacting computational efficiency and classification accuracy. The effect of clustering decisions on performance remains an area for further investigation.

\subsubsection{Detection Performance} \label{subsubsec:result_experiment_detection}
\begin{figure}[hbt!]
\centering
\includegraphics[width=\linewidth]{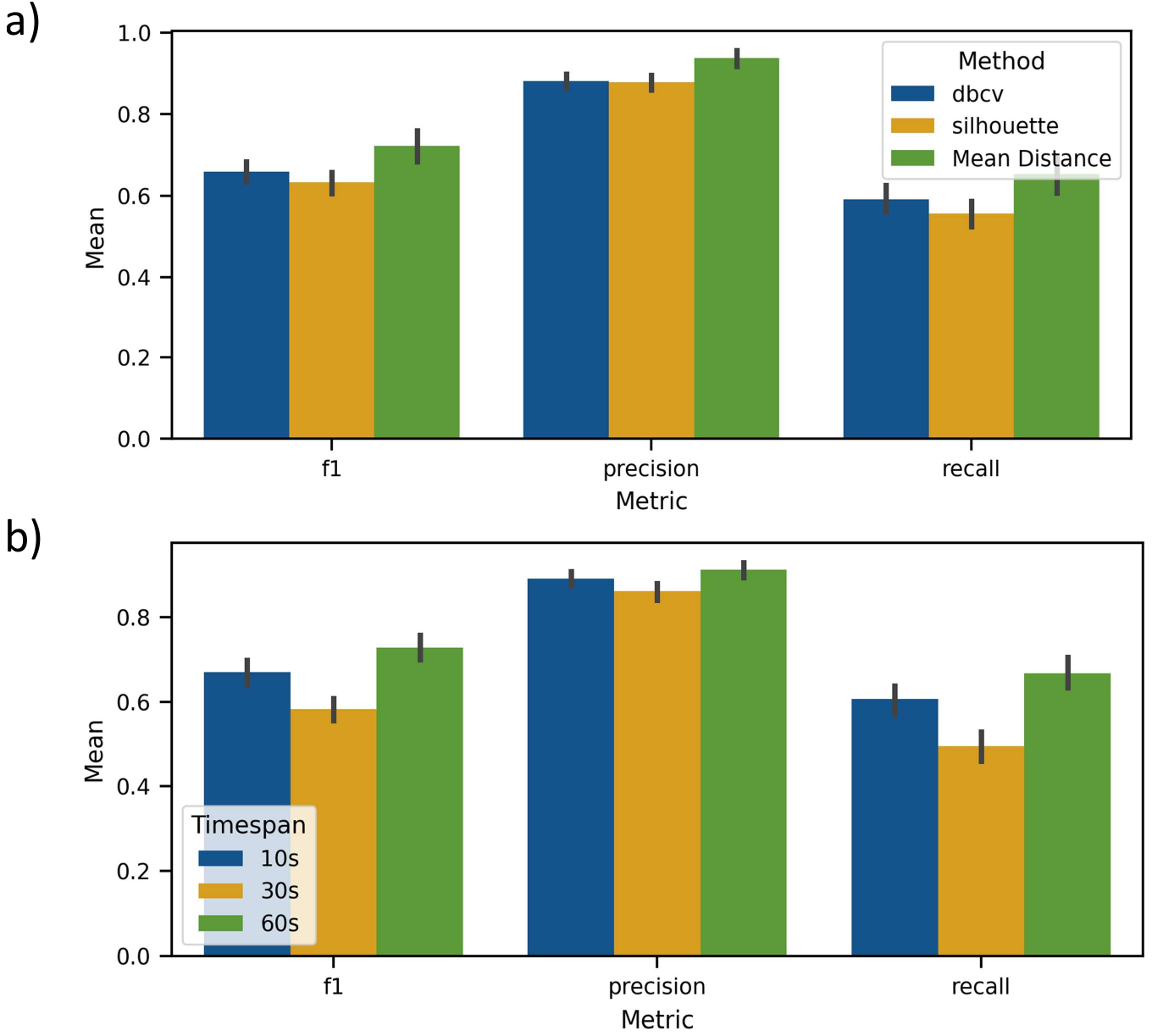}
\caption{Comparative analysis of the impact of parameter selection methods (a) and timespan (b) on performance metrics.}
\label{fig:results_detection1}
\vspace{-0.5em}
\end{figure}

\begin{figure}[hbt!]
\centering
\includegraphics[width=\linewidth]{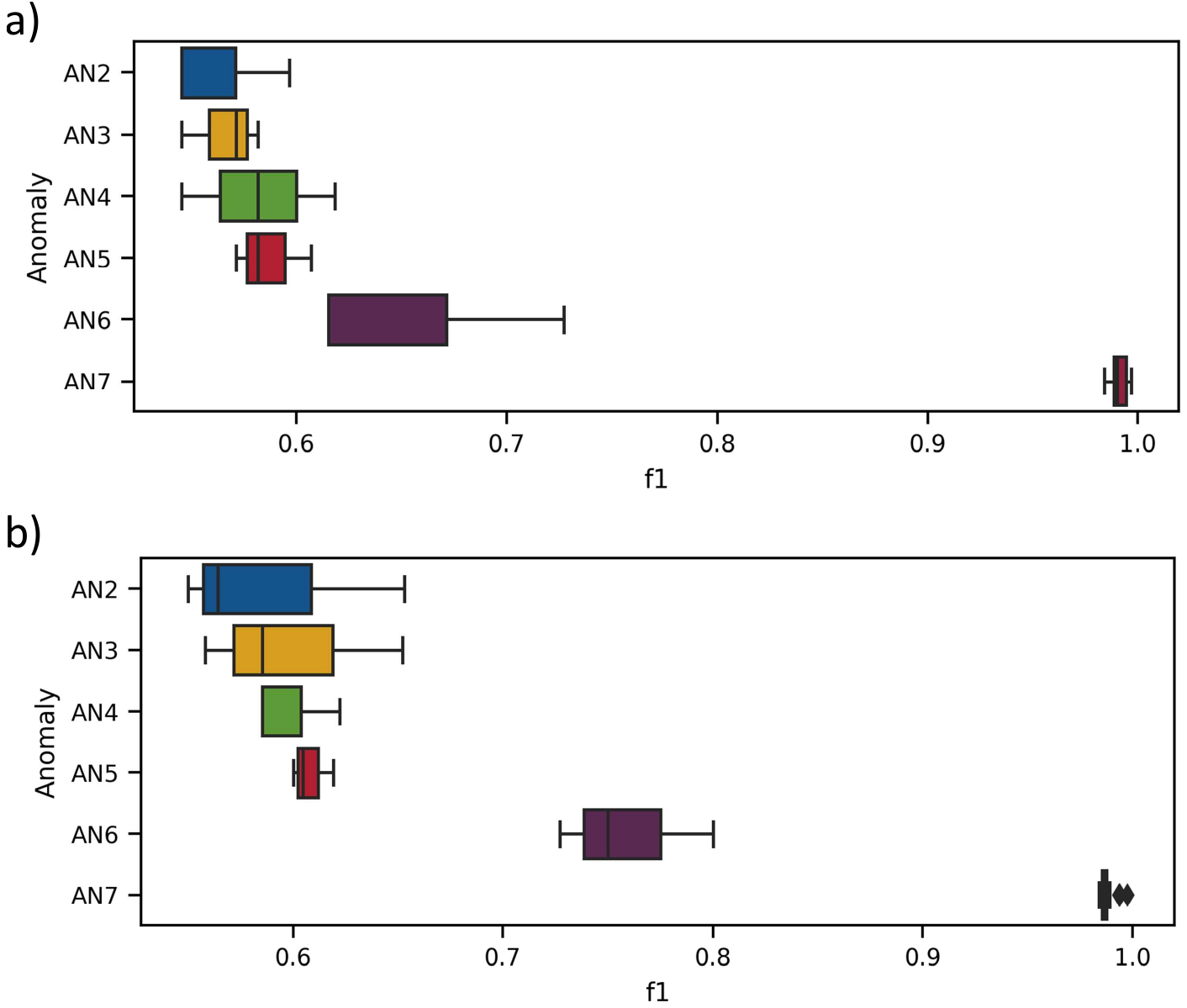}
\caption{Comparison of F1 scores by anomaly type for encrypted data (a) and unencrypted data (b) using \ac{HDBSCAN}.}
\label{fig:results_detection2}
\vspace{-1.5em}
\end{figure}

The evaluation reveal that the method proposed by ~\cite{zolotukhin2015data} excelled with the highest F1 score of 0.783, suggesting optimal precision and recall balance. This approach outperformed others in optimizing clustering for our classification model. Figures ~\ref{fig:results_detection1} illustrate this, further emphasizing the importance of the right timespan selection, with a 60-second frequently leading to better performance.

A more nuanced insight emerged from the comparative analysis of \ac{TLS} encrypted and non-encrypted datasets. Although precision metrics appeared largely consistent across both dataset types, a marked discrepancy was noted in recall and F1 scores. This discrepancy suggests that while the model is adept at accurately identifying positive instances, its ability to do so consistently varies between encrypted and non-encrypted traffic.
Further complexity in the model's performance was unveiled when analyzing different types of anomalies, as shown in Figure~\ref{fig:results_detection2}.
It revealed better detection of significant traffic changes (AN7, AN6) over subtle scenarios (AN2-AN4).
This highlights the difficulty in detecting nuanced anomalies in communication patterns within \ac{ICS} environments.
%
%

While no single algorithm or parameter set was universally optimal, some configurations consistently performed well, especially in detecting anomalies in encrypted traffic. Case studies show the successful integration of encryption-aware \ac{IDS} in IEC62531-compliant power grids, demonstrating strong potential for robust anomaly detection in secure environments, despite challenges with variable classification performance and parameter optimization.

%% file: chapter5.tex
\vspace{-0.5em}
\section{Conclusion} \label{sec:conclusion}

This paper investigates the effectiveness of an encryption-aware \ac{IDS} for unencrypted and encrypted \ac{IEC104} communication within the energy sector. Methodologically, it integrates flow metrics in slotted/windowed modes with public and digital twin-generated datasets. It tests DBSCAN and HDBSCAN clustering algorithms, using mean pairwise distance, DBCV, and Silhouette score for optimization. Findings show performance variability in anomaly detection, especially in encrypted communication. While the original approach demonstrated average success, no modifications led to consistently high-performing methods. Future research could focus on improving windowed mode, refining parameter selection, and better correlating grid anomalies with communication patterns.

